\documentstyle[12pt]{article}

\begin{document}
\title{\textbf{Primordial magnetic fields constrained from CMB anisotropies on dynamo cosmology}} \maketitle
{\sl \textbf{L.C. Garcia de Andrade}-Departamento de F\'{\i}sica
Te\'orica-IF-UERJ-RJ, Brasil\\[-0.1mm]
\vspace{0.1cm} \paragraph*{Magneto-curvature stresses could deform magnetic field lines and this would give rise to back reaction and restoring magnetic stresses [Tsagas, PRL (2001)]. Barrow et al [PRD (2008)] have shown in Friedman universe the expansion to be slow down in spatial section of negative Riemann curvatures. From Chicone et al [CMP (1997)] paper, proved that fast dynamos in compact 2D manifold implies negatively constant Riemannian curvature, here one applies the Barrow-Tsagas ideas to cosmic dynamos. Fast dynamo covariant stretching of Riemann slices of cosmic Lobachevsky plane is given. Inclusion of advection term on dynamo equations [Clarkson et al, MNRAS (2005)] is considered. In absence of advection a fast dynamo is also obtained. Viscous and restoring forces on stretching particles decrease, as magnetic rates increase. From COBE data ($\frac{{\delta}B}{B}\approx{10^{-5}}$), one computes stretching $\frac{{\delta}V^{y}}{V^{y}}=1.5\frac{{\delta}B}{B}\approx{1.5{\times}10^{-5}}$. Zeldovich et al has computed the maximum magnetic growth rate as ${\gamma}_{max}\approx{8.0{\times}10^{-1}t^{-1}}$. From COBE data one computes a lower growth rate for the magnetic field as ${\gamma}_{COBE}\approx{6.0{\times}10^{-6}t^{-1}}$, well-within Zeldovich et al estimate. Instead of the Harrison value $B\approx{t^{\frac{4}{3}}}$ one obtains the lower primordial field $B\approx{10^{-6}t}$ which yields the $B\approx{10^{-6}G}$ at the $1s$ Big Bang time.}
\newpage
\section{Introduction}
  Recently researchers at Paris observatory have discovered that faint magnetic fields around stars may be of Big Bang primordial origin. While this shows that these fields cannot be explained by the usual stellar dynamo mechanisms, it points also to the fact that a cosmological dynamo theory for the big bang primordial field is still in order. Recently according to Tsagas \cite{1}, magnetic stresses and magnetic curvature interaction introduces, stresses which slow down \cite{2} the universe expansion in open spatial sections of negative constant Riemann curvature. Growth rates of the magnetic field in the highly conducting limit. Following Arnold et al \cite{3} that the stretching of the particles are fundamental for existence of fast dynamo flows \cite{4}, and also the Chicone et al \cite{5,6}, idea that the fast dynamo operator can only be supported in two-dimensions in the case Riemannian compact manifold possesses a negative curvature, one derives here two simple solution for the Einstein´s general relativistic magnetohydrodynamics (GR-MHD) dynamo equations \cite{7} in the cases of non-ideal (slow dynamos) and ideal highly conducting limit (fast dynamos) of plasma cosmologies. In the second case, similar to the Peyrot et al \cite{8} oscillating dynamos, one obtains a restoring force dynamo in the space of constant negative curvature given by the Lobachevsky plane \cite{4}. The first case addresses the slow dynamos in the quasi-axi-symmetric absence of absent advection term. While this term is fundamental for the existence of dynamos in flat spaces, here in the cosmological relativistic setting it is shown that slow dynamos can be obtained. Of course since spacetime is locally flat this result seems to indicate that the dynamos cannot be sustained in the absence of the diffusion as in the flat space case. The important issue here is that the hyperbolic space is embedded in a (3+1)-spacetime, where 2D Riemann spaces of curvature can be embedded. Actually this result has been proved by S Carneiro \cite{9} in the framework of GR cosmology who suggested that a fast or at least slow dynamo can be embedded in cosmological spatial section of the Friedmann-Goedel metric. Recently Klebanov and Maldacena \cite{10} applied similar results to higher-dimensional anti-de Sitter cosmology (AdS). Also recently the author has shown \cite{11} that the AdS anti-de Sitter spacetime leads to a slow dynamo when flat embedding is used in cosmology. In this paper it is shown that the magnetic field growth rate can damp the restoring force in the stretching particles of the space plasma. Therefore, in principle it can bounded by COBE data.  In this paper a forcing kinematic dynamo is used as a model for magnetic field reversal in hyperbolic manifold of negative constant Riemannian curvature. Actually the advantage of using the cartesian coordinates of the Lobachevsky plane is to simplify the computation and the solution of the self-induction equation instead of using the radial and angular coordinates of the paraboloid or hyperbolic space of constant Riemann negative curvature. The study undertaken here allows us to confirm the slow of the growth of magnetic fields by Barrow and Tsagas \cite{12}, by the model of restoring force in fast dynamos in hyperbolic spaces. Besides the more physical aspects of the hyperbolic cosmic plasmas one also computes the Riemann curvature invariants to show that the apparent singularity in the space is simply a coordinate singularity like in the event horizon of a Schwarzschild black hole. Negative constant Riemann curvature has been used also by the one-component two-dimensional plasma by Fantoni and Tellez \cite{13} in the realm of electron plasmas in 2D. In the last case examined of a shearing cosmic plasma universe without expansion, the topological entropy is obtained from the volume integral of the magnetic field obtained from the MHD GR dynamo equation. Here the fold is substituted by Riemann curvature, while the stretch and shear are the usual expansion and shear of the cosmic plasma. In the non-relativistic limit, Fantoni and Tellez (J stat Phys (2008)) have used a paraboloid, on a black hole analogy, to investigate one-component two-dimensional plasmas. The paper is organized as follows: Section II presents the mathematical formalism necessary to grasp the rest of the paper. In this same section the curvature non-singular nature of the plasma is investigated, as well as the slow dynamo in the quasi-axi-symmetric approximation. In section II one also addresses the computation of the COBE limit of the particle stretching in dynamo flows in the limit of Riemann-flat sections. In section III the fast dynamo in the hyperbolic plasma is investigate and the physical interpretation of the restoring and viscous plasma flows is obtained. In this section one also computes the particle stretching from COBE data, in the high conductivity limit. In this same section the non-geodesic equation is characterized by the presence of an external force on the LHS of the geodesic equation. Section IV, one computes the topological entropy of the shearing cosmic fast dynamo in Riemannian spaces of constant negative curvature. Cosmic shear and expansion models are considered in section V and growth rate of the magnetic field is estimated from the MHD GR dynamo equation. Discussions and conclusions are presented in section VI.
\newpage
\section{COBE limits of geodesic deviation of stretched particles in cosmic dynamos}
Despite of the fact that Hilbert theorem of 1901, has shown that lower dimensional non-Euclidean manifolds of just one dimensional less could not be embedded into a three-dimensional Euclidean space, here one may circumvent this difficult problem by imagining that the 2D dimensional non-Euclidean space can be embedded in a 4-dimensional non-Euclidean spacetime, which is locally flat. Actually since spacetime is not a metrical space, this problem could be avoided simply by remembering that now the space where one is embedding to, is a non-Euclidean spaces and not in $\textbf{E}^{3}$. This section describe the basic mathematics of the Lobachevsky plane geometry for the benifit of non-mathematical inclined reader. The Lobachevsky metric is given by
\begin{equation}
ds^{2}=y^{-2}[dx^{2}+dy^{2}]
\label{1}
\end{equation}
where ${\Lambda}^{2}=(w=x+iy;y>0)$ is the hyperbolic plane in its half-upper part. Here $\sqrt{-1}=i$ is the imaginary unit of the complex plane $\textbf{C}$. Let us now compute the Ricci tensor and Riemann-Christoffel symbols of the Lobachevsky metric
\begin{equation}
R_{11}=\frac{1}{y^{2}}\label{2}
\end{equation}
\begin{equation}
R_{22}=\frac{1}{y^{2}}\label{3}
\end{equation}
\begin{equation}
R=2\label{4}
\end{equation}
\begin{equation}
{{\Gamma}^{1}}_{21}={{\Gamma}^{2}}_{22}=-\frac{1}{y}\label{5}
\end{equation}
\begin{equation}
{{\Gamma}^{2}}_{11}=\frac{1}{y}\label{6}
\end{equation}
while the Riemann curvature tensor is
\begin{equation}
{R_{1212}}=-\frac{1}{y^{4}}\label{7}
\end{equation}
The Kretschmann scalar invariant, so much used in GR to determine whether a singularity is not a true singularity or an horizon, just in Schwarzschild black hole geometry, is given by
\begin{equation}
{\cal{R}}={R_{1212}}{R^{1212}}=-1\label{8}
\end{equation}
which shows that the line $y=0$ represents a fake singularity or an event horizon of the 2D section of the universe. The process by which the particles are stretched in the cosmic flow to give rise to dynamo in the spatial section of Friedmann universe, is the geodesic equation
\begin{equation}
\frac{d^{2}J}{ds^{2}}+K(s)J=0
\label{9}
\end{equation}
whose solution for the negative curvature hyperbolic space is
\begin{equation}
J(s)=J_{0}sinh(\sqrt{-K}s)=J_{0}sinh(s)
\label{10}
\end{equation}
On this hyperbolic section one shall now solve the self-induction GR-MHD dynamo equation of Clarkson and Marklund \cite{3}, which can be expressed as
\begin{equation}
\dot{\textbf{B}}(1+\frac{5}{3}{\Theta}{\eta})-(1+\frac{2}{3}{\eta}{\Theta})\frac{2}{3}
{\Theta}\textbf{B}=
{\nabla}{\times}(\textbf{V}{\times}\textbf{B})+{\eta}{\Delta}\textbf{B}-
{\eta}<\textbf{Ric},\textbf{B}>\label{11}
\end{equation}
Note that the force-free dynamo equation yields
\begin{equation}
{\Delta}B=-curl(curl B)=-{\lambda}^{2}B
\label{12}
\end{equation}
where
\begin{equation}
curl B={\lambda}B
\label{13}
\end{equation}
is the force-free Beltrami equation. Substitution of expression (\ref{12}) into expression (\ref{11}) for the dynamo GR-MHD equation, simplifies a great deal of the equation. Together with the assumption that the magnetic and velocity fields are quasi-axi-symmetric one obtains that
\begin{equation}
curl v{\times}B=0
\label{14}
\end{equation}
Substitution of these expressions
\begin{equation}
\dot{\textbf{B}}(1+\frac{5}{3}{\eta}{\Theta})-[(1+{\eta}\frac{2}{3}{\Theta})\frac{2}{3}
{\Theta}-{\lambda}^{2}{\eta}]\textbf{B}=-{\eta}<\textbf{Ric},\textbf{B}>\label{15}
\end{equation}
Here the Ricci curvature term is
\begin{equation}
<\textbf{Ric},\textbf{B}>={R^{i}}_{j}B^{j}\label{16}
\end{equation}
where (i,j=1,2,3) and $R_{ij}$ are the components of the two or three-dimensional Ricci curvature tensor. Thus in components this equation reduces to
\begin{equation}
[{\gamma}(1+\frac{5}{3}{\eta}{\Theta})+(1+{\eta}\frac{2}{3}{\Theta})\frac{2}{3}
{\Theta}-{\lambda}^{2}{\eta}]B^{i}=-{\eta}{R^{i}}_{j}B^{j}\label{17}
\end{equation}
where ${\gamma}$ is the rate of the amplification of the magnetic field and ${\Theta}=\frac{\dot{V}}{V}$. Another important feature is that to simplify the complicated Clarkson-Marklund dynamo GR equation one has assumed that the expansion of the universe is slow $\dot{{\Theta}}\approx{0}$ and the expansion is linear or ${\Theta}^{2}\approx{0}$. With these simplifications at hand one may simply demonstrate the constraints imposed by the fast dynamo action on the cosmic plasma model as
\begin{equation}
{\gamma}=-\frac{[(1+{\eta})\frac{2}{3}
{\Theta}-{\lambda}^{2}{\eta}]}{(1+\frac{5}{3}{\eta}{\Theta})}\label{18}
\end{equation}
By making use of the fast dynamo limit criteria \cite{16}
\begin{equation}
lim_{{\eta}\rightarrow{0}}\textbf{Re}{\gamma}(\eta)>0
\label{19}
\end{equation}
one obtains
\begin{equation}
lim_{{\eta}\rightarrow{0}}\textbf{Re}{\gamma}(\eta)=-\frac{2}{3}{\Theta}
\label{20}
\end{equation}
Here $\textbf{Re}$ represents the real part of the growth rate scalar ${\gamma}$. This shows that the fast dynamo shall be supported, only when ${\Theta}$ is negative or in physical terms when the universe undergoes contracting phases. It is clear that when nonlinear expansion terms are allowed this situation shall not be modified, but then acceleration terms would have to be taken into account and the dynamo equation complicates considerable. Thus
\begin{equation}
|\frac{{\delta}B}{B}|=\frac{2}{3}\frac{\dot{{\delta}V^{y}}}{V^{y}}\label{21}
\end{equation}
which from COBE data
\begin{equation}
\frac{{{\delta}B}}{B}={10^{-5}}
\label{22}
\end{equation}
In the highly conducting $({\eta}=0)$
ideal plasma cosmologies, the stretching of particles in the flow
\begin{equation}
\frac{{\delta}{V^{y}}}{V^{y}}\approx{1.5{\times}10^{-5}}
\label{23}
\end{equation}
In the next section advection terms is considered and the fast dynamo constraints are modified.
\newpage
\section{Restoring force and advection in fast dynamos}
In this section one shall address the problem of solving the Clarkson-Marklund equation in full detail in the case of force-free MHD dynamo and present the constraints under which the dynamo becomes fast. Existence of fast dynamos in negatively constant curved 2D-Riemann space has been proved by Chicone et al \cite{5} and here one simply applies this result to cosmic plasma. Contrary to the previous example here one considers the advection terms in terms of the universe expansion ${\Theta}$. The fact that one is not considering vorticity and shear of the cosmological model is that actual universe can be perfectly approximated by the expansion, shear-free and vorticity-free Friedmann model. Friedmann model is also suitable to two and three-dimensional negative curved spatial section as has been recently shown by Carneiro \cite{8}. Actually restoring forces seems  damping the expansion of the universe which in turn makes it to contract favouring fast dynamo action. The main ingredient in this section is the advection term, which can be expressed as
\begin{equation}
[curl(\textbf{V}{\times}\textbf{B})]^{i}=\frac{1}{\sqrt{g}}[V^{i}(divB)+{\Theta}B^{i}\sqrt{g}+
\sqrt{g}B^{k}V^{l}{{\Gamma}^{i}}_{kl}-(divV)B^{i}+\sqrt{g}V^{m}{\partial}_{m}B^{i}]
\label{24}
\end{equation}
Due to the solenoidal character of the flow and magnetic fields, this expression reduces to
\begin{equation}
[curl(\textbf{V}{\times}\textbf{B})]^{i}=\frac{1}{\sqrt{g}}[{\Theta}B^{i}\sqrt{g}+
\sqrt{g}B^{k}V^{l}{{\Gamma}^{i}}_{kl}+\sqrt{g}V^{m}{\partial}_{m}B^{i}]
\label{25}
\end{equation}
The last term on the RHS of this equation is the stretching term in a non-covariant form and ${{\Gamma}^{i}}_{jk}$ is the Riemann-Christoffel symbol. Here g is the determinant of the Lobachevski metric, given by
\begin{equation}
{\sqrt{g}}=y^{-2}
\label{26}
\end{equation}
Force-free fields in the fast dynamo of cosmic Riemannian plasma slice possess the following self-induction equation
\begin{equation}
[{\gamma}(1+\frac{5}{3}{\eta}{\Theta})+(1+{\eta}\frac{2}{3}{\Theta})\frac{2}{3}
{\Theta}-{\lambda}^{2}{\eta}]B^{i}+{\eta}{R^{i}}_{j}B^{j}=[{\Theta}-\textbf{V}.{\nabla}]B^{i}
+{{\Gamma}^{i}}_{kl}V^{l}B^{k}+{\eta}{\chi}^{i}\label{27}
\end{equation}
where ${\chi}$ is a very long expression in terms of rotation, shear and expansion which can be found in the original paper by Clarkson and Marklund. In the high conductivity limit (${\eta}\rightarrow{0}$) this expression reduces to
\begin{equation}
[{\gamma}-\frac{1}{3}{\Theta}+\textbf{V}.{\nabla}]B^{i}=
{{\Gamma}^{i}}_{kl}V^{l}B^{k}\label{28}
\end{equation}
Note from Riemannian geometry that the RHS term is nothing but $B^{k}D_{k}V^{i}$ where $D_{i}$ is the Riemannian covariant derivative operator. This shows that this term is the covariant stretching term of the Riemannian dynamo flow. Note that the growth rate for dynamo action ${\gamma}$ can be expressed as
\begin{equation}
[{\gamma}-\frac{1}{3}{\Theta}+V^{y}{\partial}_{y}]B^{i}=-B^{k}V^{l}{{\Gamma}^{i}}_{kl}\label{29}
\end{equation}
Substitution of the above Riemann-Christoffel symbols one obtains
\begin{equation}
[{\gamma}-\frac{1}{3}{\Theta}+V^{y}{\partial}_{y}]B^{1}=-B^{1}V^{2}\label{30}
\end{equation}
\begin{equation}
{\gamma}=V_{0}-\frac{1}{3}{\Theta}=V_{0}-\frac{{\delta}V^{y}}{V^{y}}\label{31}
\end{equation}
From this last expression one notices that a contracting phase of the universe given by ${\Theta}<0$ implies that the open universe dynamo action can be enhanced as in gravitational collapse. It also shows that in this universe the stretching of the particles can contribute to dynamo action, a well-known fact in dynamo theory. By substituting the expression $\dot{V^{y}}=F^{y}$ in this equation one obtains
\begin{equation}
F^{y}=(V_{0}-{\gamma}){V^{y}}\label{32}
\end{equation}
which is a viscous cosmic flow. Now let us consider the non-geodesic equation
\begin{equation}
\frac{dV^{y}}{dt}+{{\Gamma}^{y}}_{yy}(V^{y})^{2}=F^{y}(y)
\label{33}
\end{equation}
which yields in the Lobachevsky space
\begin{equation}
\frac{dV^{y}}{dt}-\frac{1}{{y}}(V^{y})^{2}=F^{y}(y)
\label{34}
\end{equation}
From the solenoidal equation, one obtains $V^{y}=V_{0}y^{2}$ which upon substituting into the non-geodesic equation, with external force $F^{y}=\dot{V^{y}}$ yields the restoring force
\begin{equation}
F^{y}=-k{y}
\label{35}
\end{equation}
where $k:=V_{0}(V_{0}-2)$. This shows that the strong velocity flows, such as in turbulent flows, may induce a restoring force. In the next section one shall show that cosmic shear can also contribute to cosmic dynamo action in hyperbolic spatial sections of the universe, even in the absence of expansion.
\section{Dynamo action and primordial magnetic fields from COBE data}
Earlier efforts in the direction of the construction of a cosmological dynamo theory were done by Zeldovich et al \cite{14} who estimated dynamo magnetic field growth rate in the cosmic scale as ${\gamma}_{max}\approx{8.0{\times}10^{-1}t^{-1}}$ for a low  magnetic field of $B\approx{t^{\frac{4}{5}}}$. In this section from the expression (\ref{11}) in the ideal resistivity-free, highly conductive plasma cosmology, it is easy to show that the magnetic field evolution would be
\begin{equation}
\frac{\dot{B}}{B}={\gamma}=\frac{2}{3}{\Theta}=\frac{\dot{V}}{V}
\label{36}
\end{equation}
From this expression an expanding universe, such as the one with conformally flat Riemann spatial sections considered here, yields a relation between the stretching and the COBE data as seen above. From a simple analysis of the above formula yields the following growth rate
\begin{equation}
{\gamma}=\frac{2}{3}{\Theta}=\frac{2}{3}<\frac{{\delta}B}{B}>t^{-1}\approx{6.0{\times}10^{-6}t^{-1}}
\label{37}
\end{equation}
Note that the case of the Zeldovich et al yields a magnetic field yields $B\approx{t^{\frac{4}{3}}}$ while here one obtains that $B\approx{10^{-6}t}$ which at $1s$ of Big Bang is the primordial well-known result of the order of $B\approx{10^{-6}G}$. When takes into account shearing effects in the Clarkson-Marklund equation one obtains
\begin{equation}
{\gamma}=[\frac{2}{3}{\Theta}+{\sigma}]
\label{38}
\end{equation}
which in the absence of covariant stretch shows that the growth rate vanishes. Here ${\sigma}$ represents the trace of the shear tensor. This can be simply shown if one considers the expression for the shear tensor ${\sigma}_{ij}$ as
\begin{equation}
{\sigma}_{ij}={h^{m}}_{[i}{h^{n}}_{j]}D_{m}V_{n}-\frac{1}{3}{\Theta}h_{ij}
\label{39}
\end{equation}
Here 2D metric $h_{ij}$ is given by
\begin{equation}
{h}_{ij}=g_{ij}+V_{i}V_{j}
\label{40}
\end{equation}
Thus in the absence of the flow stretching term with the covariant derivative operator $D_{m}$ one obtains
\begin{equation}
{\sigma}=Tr{{\sigma}_{ij}}=-\frac{2}{3}{\Theta}
\label{41}
\end{equation}
Substitution of this expression into (\ref{38}) yields the vanishing of the growth rate. This actually can characterize a slow dynamo action.
\section{Conclusions}
A toy model of a spatial hyperbolic section of a cosmological model as a fabric of spacetime dynamos in Lobachevsky plane is considered. In this model the stretching of particles in the universe as a model for fast cosmic dynamos in plasmas is given. Restoring and viscous forces in the model are obtained from the geodesic equation. Particle stretching, a fundamental ingredient for fast dynamos is estimated from COBE data of background radiation. In these spaces of negative constant Riemannian curvature known as Lobachevsky hyperbolic planes, the coordinate and singularity is lifted by computing the curvature invariants and the model is actually non-singular spatially. Earlier Garcia de Andrade \cite{15} has investigated the fast dynamo by stretching of plasma flows. This paper can be extended to ideal dynamo plasmas cosmologies in near future. COBE bounds in the GR MHD dynamo cosmology seems to favour the model obtained by Clarkson and Marklund to explain primordial magnetic fields by dynamo cosmological mechanisms.
\section{Acknowledgements}
I am very much indebt to Dmitry Sokoloff, Hermann Mosquera and Yu Latushkin, for reading the manuscript and for helpful discussions on the subject of this work. I appreciate financial  supports from UERJ and CNPq.

  \end{document}